\documentclass[twocolumn]{agujournal2019}
\usepackage{url} %this package should fix any errors with URLs in refs.
\usepackage{lineno}
\usepackage[inline]{trackchanges} %for better track changes. finalnew option will compile document with changes incorporated.
\usepackage{soul}

\begin{document}

\title{A Merged Search-Coil and Fluxgate Magnetometer Data Product for Parker Solar Probe FIELDS}

\authors{Trevor A. Bowen\affil{1}, Stuart D. Bale\affil{1,2}, John W. Bonnell\affil{1}, Thierry Dudok de Wit\affil{3}, Keith Goetz\affil{4}, Katherine Goodrich\affil{1}, Jacob Gruesbeck\affil{5}, Peter R. Harvey\affil{1}, Guillaume Jannet\affil{3}, Andriy Koval\affil{6,7}, Robert J. MacDowall\affil{5}, David M. Malaspina\affil{8}, Marc Pulupa\affil{1},  Claire Revillet\affil{3}, David Sheppard\affil{5}, Adam Szabo\affil{7}}
%
%\affiliation{1}{Physics Department and Space Sciences Laboratory,
%University of California, 7 Gauss Way, Berkeley, CA 94720, USA }%
%\affiliation{2}{CNRS}%
%\affiliation{3}{Goddard}%

%
%\altaffiltext{1}{Space Sciences Laboratory, University of California, Berkeley, CA, USA}
%\altaffiltext{2}{Physics Department, University of California, Berkeley, CA, USA}
%\altaffiltext{3}{School of Physics and Astronomy, University of Minnesota, Minneapolis, MN, USA}
\correspondingauthor{Trevor A. Bowen}{tbowen@berkeley.edu}

\affiliation{1}{Space Sciences Laboratory, University of California, Berkeley, CA 94720-7450, USA}
\affiliation{2}{Physics Department, University of California, Berkeley, CA 94720-7300, USA} 
\affiliation{3}{LPC2E, CNRS and University of Orl\'{e}ans, Orleans, France}
\affiliation{4}{School of Physics and Astronomy, University of Minnesota, Minneapolis, MN 55455}
\affiliation{5}{Planetary Magnetospheres Laboratory, NASA Goddard Space Flight Center, Greenbelt, MD 20771, USA}
\affiliation{6}{Goddard Planetary Heliophysics Institute, University of Maryland, Baltimore County, Baltimore, MD 21250, USA}
\affiliation{7}{Heliospheric Physics Laboratory, NASA Goddard Space Flight Center, Greenbelt, MD 20771, USA}
\affiliation{8}{Laboratory for Atmospheric and Space Physics, University of Colorado, Boulder, CO 80303, USA}

%14Code 695, NASA Goddard Space Flight Center, Greenbelt, MD 20771, USA
%15Goddard Planetary Heliophysics Institute, University of Maryland Baltimore County, Baltimore, MD 21250, USA
%16Code 672, NASA Goddard Space Flight Center, Greenbelt, MD 20771, USA
%17Department of Physics and Astronomy, University of Iowa, Iowa City, IA 52242, USA
%18Climate and Space Sciences and Engineering, University of Michigan, Ann Arbor, MI 48109, USA
%19Heliophysics Division, NASA Goddard Space Flight Center, Greenbelt, MD  20771, USA
%20University of Applied Sciences and Arts Northwestern Switzerland, 5210 Windisch, Switzerland 
%21LESIA, Observatoire de Paris, Universite PSL, CNRS, Sorbonne Universite, 92195 Meudon,France
%22Department of Astrophysical Sciences, Princeton University, Princeton, NJ 08540, USA
%23Department of Earth, Planetary, and Space Sciences, University of California, Los Angeles, USA

\begin{abstract}

NASA's Parker Solar Probe (PSP) mission is currently investigating the local plasma environment of the inner-heliosphere ($< $0.25$R_\odot$) using both {\em{in-situ}} and remote sensing instrumentation. Connecting signatures of microphysical particle heating and acceleration processes to macro-scale heliospheric structure requires sensitive measurements of electromagnetic fields over a large range of physical scales. The FIELDS instrument, which provides PSP with {\em{in-situ}} measurements of electromagnetic fields of the inner heliosphere and corona, includes a set of three vector magnetometers: two fluxgate magnetometers (MAGs), and a single inductively coupled search-coil magnetometer (SCM). Together, the three FIELDS magnetometers enable measurements of the local magnetic field with a bandwidth ranging from DC to 1 MHz. This manuscript reports on the development of a merged data set combining SCM and MAG (SCaM) measurements, enabling the highest fidelity data product with an optimal signal to noise ratio. On-ground characterization tests of complex instrumental responses and noise floors are discussed as well as application to the in-flight calibration of FIELDS data.  The algorithm used on PSP/FIELDS to merge waveform observations from multiple sensors with optimal signal to noise characteristics is presented. In-flight analysis of calibrations and merging algorithm performance demonstrates a timing accuracy to well within the survey rate sample period of $\sim340 \mu s$. \end{abstract}

\newcommand{\bvec}[1]{{\ensuremath{\bf{#1}}}}

%\chapter{Merging Search Coil and Fluxgate Magnetic Field Data for Parker Solar Probe FIELDS}
%\label{chap7}
\section{Introduction}
The {\em{in-situ}} measurements of coronal heating, solar wind acceleration, and energetic particle transport made by NASA's Parker Solar Probe (PSP) will likely answer many fundamental questions relating to the heliosphere and astrophysical plasmas \cite{Fox2016}. The FIELDS instrument on PSP provides {\em{in-situ}} measurements of the electric and magnetic fields \cite{Bale2016} required to achieve the principal objectives. The measurements made by FIELDS are complemented by {\em{in-situ}} measurements of the solar wind and coronal plasma through the Solar Wind Electrons Alphas and Protons \cite<SWEAP,>{Kasper2016} investigation; energetic particles through the Integrated Science Investigation of the Sun \cite<IS$\bigodot$IS,>{McComas2016}; as well as white light images from the Wide-Field Imager for Solar Probe \cite<WISPR,>{Vourlidas2016}.

Accomplishing the scientific objectives of PSP requires {\em{in-situ}} observations of magnetic and electric fields over a wide bandwidth and large dynamic range. FIELDS measures electric field and potential fluctuations ranging from DC-19.2 MHz with a diverse combination of survey mode waveform, burst mode waveform, and spectral observations made by the FIELDS Radio Frequency Spectrometer \cite<RFS,>{Pulupa2017}, Digital Fields Board \cite<DFB,>{Malaspina2016}, and Time Domain Sampler (TDS) \cite{Bale2016}. Magnetic field instrumentation consists of a suite of three magnetometers, two vector fluxgate magnetometers (MAGs) and a single vector search-coil magnetometer (SCM) located on boom extending behind the spacecraft and within the umbra of the PSP thermal protection system (TPS).

The two fluxgate magnetometers, built at Goddard Space Flight Center (GSFC), provide vector measurements of DC and low frequency magnetic fields with a maximum survey sample (Sa) rate of $f_{svy}^{max}$=292.969 Sa/s. These low frequency measurements are accompanied by measurements from the FIELDS SCM low frequency (LF) windings sensitive from $\approx$3 Hz-20 kHz. The SCM sensor $x$-axis additionally contains a second mid-frequency (MF) winding sensitive from $\approx$10 kHz-1 MHz. During the perihelion encounters the SCM is continuously sampled by the DFB typically at  $f_{svy}$=292.969 Sa/s, though a maximum rate of 18,750 Sa/s is theoretically possible \cite{Malaspina2016}. When possible, data is additionally acquired during the aphelion cruise phases at reduced sample rates in accordance with allowed telemetry and spacecraft operations. Burst mode waveforms for the LF or MF are sampled up to a maximum rate of 150 kSa/s by the DFB; the MF channel can be sampled at 1.92 MSa/s by the TDS and incorporated in the spectral products generated by the RFS. Generally, the DFB, RFS, and TDS, are highly configurable and generate a variety of waveform and cross and auto-spectral matrix data products at various cadences.

Observational signatures of physical processes occurring in astrophysical plasmas, such as the solar wind and corona, are commonly sensitive to properties of the mean magnetic field: e.g. electromagnetic wave vector polarizations \cite{PodestaGary2011a,He2011}; magnetic compressibility \cite{Bale2009,Alexandrova2013}; variance and wave-vector anisotropies \cite{Horbury2008,Chen2010,Horbury2012}; MHD Els\"{a}sser and Poynting flux \cite{Balogh1999,McManus2019}; helicical signatures of turbulence at kinetic scales \cite{Leamon1998,HowesQuataert2010,Woodham2018} magnetic reconnection \cite{Phan2018}. Though microphysical processes in the solar wind are sensitive to the low frequency mean field behavior, they frequently occur on fast time scales with small amplitude magnetic signatures. Accomplishing the scientific objectives of PSP thus inherently requires {\em{in-situ}} observations of magnetic fields over large bandwidths and dynamic ranges. Through combining both fluxgate and search-coil measurements, FIELDS is capable of observing the {\em{in-situ}} magnetic field of the inner heliosphere with a bandwidth from DC-1 MHz and 115 dB of dynamic range \cite{Bale2016}.

Recently, missions with multi-sensor observational suites have moved towards merged data products composed of synchronous measurements from multiple instruments, combined to employ optimal qualities of the separate sensors. \citeA{Alexandrova2004} used a discrete wavelet transform to merge {\em{CLUSTER}} fluxgate and search-coil data to study waves downstream of a shock \cite{ClusterMAG,ClusterSTAFF}. \citeA{Chen2010} use a similar method to study the variance anisotropy of kinetic turbulence in the sub-proton range with {\em{CLUSTER}}. Additionally, \citeA{Kiyani2009} perform measurements of scaling functions of the transition from the MHD to kinetic turbulence through continuous analysis of a combination of CLUSTER fluxgate and search-coil magnetometer observations. These previous efforts to combine fluxgate and search-coil data, though made in-flight and without specific optimization towards instrument design and operation, have proved the utility and applicability of multi-rate data fusion methods in space plasma physics \cite{Hall1992}. More recently, programmatic efforts to perform quantitative end-to-end testing on the Multiscale Magnetospheric Mission (MMS) search-coil and digital and analog fluxgate magnetometer have enabled the development of an optimized merged data set with automated calibration pipeline\cite{LeContel2016,Russell2016,Torbert2016,Fischer2016}.

The FIELDS SCM and MAG share a master clock and were designed with partially overlapping bandwidths, enabling the combination of individual sensors into a single merged dataset. This manuscript outlines the process used to produce survey data product using merged SCM  and MAG (SCaM) measurements with a spectral composition that retains an optimal signal to noise ratio. Section \ref{sec:cal} provides an overview of preflight ground testing and instrument characterization, as well as inflight calibration routines. Section \ref{sec:merging} presents the algorithm used to combine the FIELDS magnetometer data into a merged product with optimal signal to noise characteristics. Section \ref{sec:inflight} provides a summary of our in-flight verification of the calibration and a quantitative analysis of the performance of the merged survey data, and discusses merging survey rate data with DFB burst data \cite{Malaspina2016}.

\section{FIELDS Calibration}
\label{sec:cal}
\subsection{Fluxgate Magnetometers (MAG)}
The two fluxgate magnetometers, designed and fabricated at NASA/GSFC, measure DC and low frequency fluctuating magnetic fields. They are placed 1.9 and 2.7 meters from the spacecraft and are respectively referred to as the inboard (MAGi) and outboard (MAGo) sensors. The heritage of the PSP/FIELDS MAGs dates to the 1960's NASA {\em{Explorer 33}} mission \cite{Ness1971,Acuna1974,Acuna2002}. Many iterations of the instrument currently operate on both NASA heliophysics and planetary science missions \cite{Lepping1995,Acuna2002, Acuna2008, Kletzing2013, Connerney2015, Connerney2017}. The PSP fluxgate magnetometers have a maximum survey mode sample rate of $f_{svy}^{max}=292.969$ Sa/s. The MAG data is typically downsampled by factors of two with anti-aliasing performed with a Bartlett filter. Generally, MAGi is run at a lower sample rate in order to meet telemetry constraints imposed by the spacecraft. The lower cadence measurements still allow for diagnosis of magnetic noise associated with spacecraft generated magnetic fields. The primary science instrument for the DC magnetic fields is MAGo, which is less sensitive to spacecraft generated fields due to its positioning on the spacecraft boom.

The complex transfer functions associated with the MAGs, shown in Figure \ref{merging_resp}(a), are dominated by a single pole low-pass Butterworth filter used for anti-aliasing purposes tuned to -3 dB at the max sample rate Nyquist frequency ($f_{svy}^{Ny}\approx$146.5Hz) \cite{Acuna2008,Connerney2015,Connerney2017}. Due to the low-pass characteristics of the Butterworth transfer function, the MAGs are sensitive to the DC magnetic fields associated with the spacecraft \cite{Ness1970,Ness1971,Belcher1973}. Typically, the minimization of such fields is performed through magnetic control programs \cite{Ness1970,Musmann1988}. For PSP a strict magnetic cleanliness program was followed during design and development. Once in space, driven spacecraft maneuvers are used to establish the magnetometer zero offsets relative to the ambient field e.g. \cite{Acuna2002,Connerney2015}. However, similar results can be accomplished without controlled maneuvers through statistical analysis of non-compressive Alfv\'{e}nic rotations in the solar wind \cite{Belcher1973,Leinweber2008}. Several multi-sensor techniques to determine sensor zeros have been developed using gradiometric principles, e.g. \cite{Ness1971} and comparison with scalar magnitude instruments, e.g. \cite{Olsen2003,Primdahl2006}. Additionally solar wind electron beams, sensitive to the mean field direction, can be used in calibrating fluxgate offsets \cite{Plaschke2014,Connerney2015}. 

%For spinning spacecraft, the spin period allows for identification of the static spacecraft zeros within the spin plane, while offsets along the spacecraft axis are determined using ,\cite{Farreletal1995,Plaske2014}.

\begin{figure}\centering
\includegraphics[width=4.5in]{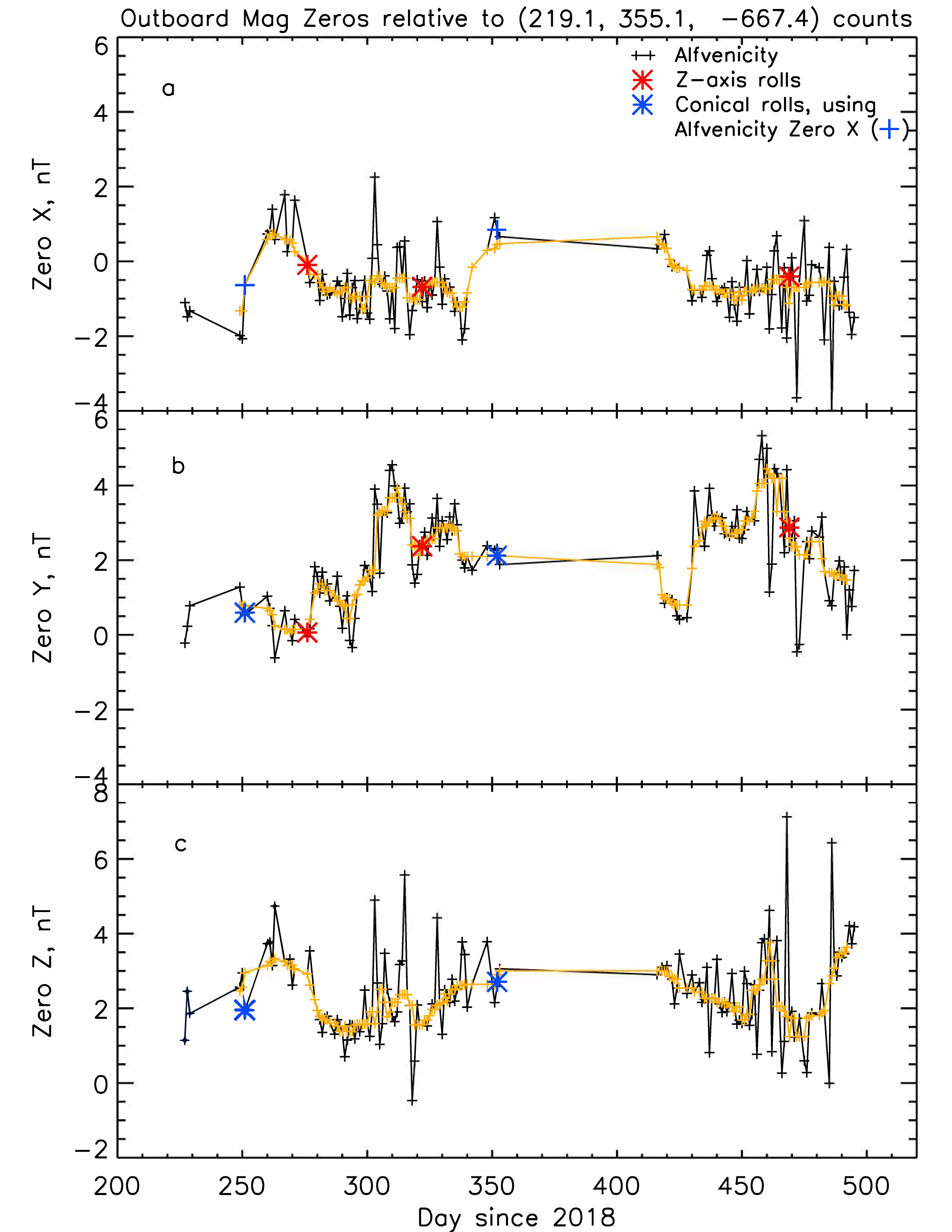}
\caption{Magnetic field offsets for FIELDS/MAGo $x$,$y$,$ and z$ axes (panels a, b, and c)  over first two encounters of PSP (respectively offset by 219.1, 355.1, -667.4 sensor counts for demonstrative purposes). Data points (+) correspond to estimates from Alfv\'enic rotations of the solar wind; a seven point median filter used in processing the FIELDS data (orange). Offsets determined from spacecraft rotations are shown in red (sun pointed rolls around the spacecraft $z$-axis) and blue (conical rolls performed off sensor $z$-axis). The blue +  corresponds to the MAGo $x$ offset determined from Alfvenic at the time of the conical roles.}\label{fig:rolls}
\end{figure}

For PSP, the attitude and pointing requirements of the spacecraft preclude the use of controlled maneuvers during perihelion, spacecraft rolls (both sun-pointed and conical rotations) are thus performed before and after each perihelion encounter to establish zero levels of the spacecraft magnetic field. In between such controlled rotations, measurements of Alfv\'{e}nic rotations of the solar wind magnetic field have been implemented to track variations in the spacecraft field, allowing for an estimate of zero levels during each perihelion encounter, when controlled maneuvers cannot be performed \cite{Leinweber2008}. Figure \ref{fig:rolls} shows the offsets for MAGo over the first encounter computed using both spacecraft rolls, and higher rate estimations from Alfv\'{e}nic rotations. Spacecraft housekeeping and engineering data are currently under analysis in order to model the effects of variations in the solar panel array, which change orientation over the course of the orbit, on the measured DC offsets. Gradiometric techniques are in development to further verify and monitor offset drifts during each perihelion encounter. In addition to the removal of spacecraft offsets, the vector axes of each MAG are orthogonalized using an alignment matrix determined during pre-flight testing. The alignment matrix, is determined through the process documented in \citeA{Acuna1981} and \citeA{Connerney2017} and verified by methods outlined in \citeA{Risbo2003}.

In addition to the removal of spacecraft offsets and sensor orthogonalization, the merged data set corrects gain and phase shifts associated with the analog Butterworth filter using a convolution of the MAG output with the linear time-invariant inverse filter response. When appropriate, the digital Bartlett filters used to downsample MAG measurements are additionally inverted. Though the phase shifts associated with these low pass filters are quite small (e.g. Figure \ref{merging_resp}) over the range of frequencies observed by both the MAG and SCM, the mis-alignment of the relative phase between the MAG and SCM leads to undesirable artifacts in the merged data product. Correction for complex instrumental response (i.e. filter gain and phase) is not implemented in the un-merged calibrated data.

%
%The MAGSAT method is based upon a heritage procedure for vector magnetometer absolute sensor alignment determination and is described by M.H. Acuña [10]. This is a modified version of the algorithm presented by McPherron and Snare [11] representing a process by which the absolute alignment of the three magnetic axes of the SPF MAG are determined relative to an external, fixed reference coordinate system. In the MAGSAT method, three sensor positions are used, aligned to the facility coil system via the facility first-order theodolites and an optical cube bonded to the SPF MAG sensor. This magneto-optical method obviates the need to precisely determine the absolute orientation of the applied field via the geometry of the calibration coils and can produce results with RMS variations of less than 10 arc-seconds [4]. This level of accuracy greatly exceeds the FIELDS requirements.
%

\subsection{Search-Coil Magnetometer (SCM)}

\begin{figure}\centering
\includegraphics{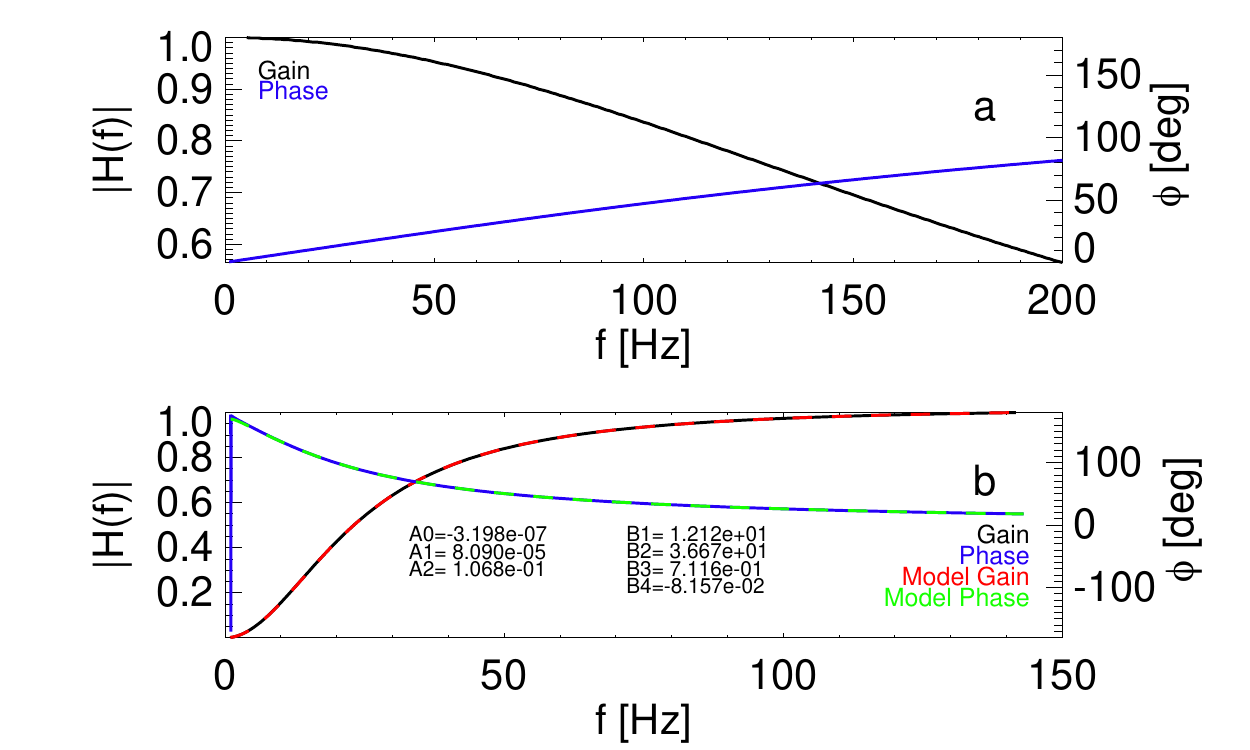}
\caption{(a) MAGo frequency response is dominated by single pole Butterworth filter response tuned to -3 dB at the survey mode Nyquist frequency ($f_{svy}^{max}/2$). (b) SCM frequency response determined from a spectral analyzer. A 4-pole, 2-zero fit analytical fit is performed to the empirically determined function.}
\label{merging_resp}
\end{figure}

The PSP/FIELDS search-coil magnetometer was designed at Laboratoire de Physique et Chimie de l'Environnement et de l'Espace (LPC2E) in Orl\'{e}ans, France, and is nearly identical to the search-coils which have been manufactured and delivered for the Solar Orbiter and TARANIS missions \cite{SeranFergeau2005,Maksimovic2019}. The SCM, consisting of three mutually orthogonal inductive coils mounted on the end of the spacecraft magnetometer boom, is tailored to study the magnetic fields of the inner heliosphere \cite{Bale2016} from 10-20 kHz. Additionally, one of the coils (sensor $x$-axis) has a secondary winding with bandwidth of 10 kHz-1 MHz. 

The survey mode waveform data, typically captured at a rate of $f_{svy}^{max}=292.969$ Sa/s is sampled and processed by the FIELDS DFB \cite{Malaspina2016}. In addition to continuous survey mode waveform data, the DFB provides a highly configurable set of operational modes which can be modified in flight to generate a diverse set of burst waveform and spectral data products. In addition to sampling the SCM output, the DFB is designed to inject a programmable calibration signal into the SCM. The response of the SCM to the injected stimulus is captured by the DFB as well as the TDS and the instrumental transfer function can accordingly be determined in-flight.

Figure \ref{merging_resp}(b) shows the gain and phase characteristics of the FIELDS SCM $x$-axis which were determined empirically on ground using a spectral analyzer. A complex rational function with 4-poles and 2-zeros:  \begin{equation}
R(i\omega) =\frac{A_0+i\omega A_1}{(B_0-B_2\omega^2)+i\omega(B_1-B_3\omega^2)}
\end{equation}
is fit to the the response using the least square estimation techniques developed by \citeA{Levy1959}.

The inductive nature of the SCM leads to strong gain and phase shifts in the instrumental response function which must be compensated to obtain an estimate of the observed magnetic field in physical units. During pre-launch integration and testing different methods to invert the SCM frequency response were explored: convolution kernel methods and a windowed fast Fourier transform (FFT) algorithm similar to techniques used in \citeA{LeContel2008} and \citeA{Robert2014}. Preflight Monte-Carlo simulations testing on synthetic data suggested that convolution in the time domain generated fewer spectral artifacts in the calibrated time series than a windowed FFT algorithm. Compensation filters are developed using the inverse FFT of the response function on an abscissae of 2048 frequencies, corresponding to a 2048 tap (all zero) linear time invariant (LTI) finite impulse response (FIR) filter \cite{OppenheimSchafer}. The filters are non-causal, such that real time merging of data (e.g. on the spacecraft) is not possible; the future development of causal FIR filters for on-board merging presents an opportunity to increase scientific returns from telemetry limited missions.

\section{Merging}
\label{sec:merging}

Many algorithms for merging data from multiple sensors, occasionally referred to as data fusion, were initially developed in the context of radio system engineering as a method to optimize signal to noise ratios and correct for signal loss due to stochastic fluctuations impacting transmission, \cite{Kahn1954,Brennan1959}. Recent research has demonstrated the applicability of data fusion in merging magnetic field measurements from multiple sensors onboard a single spacecraft: \cite{Alexandrova2004,Kiyani2009,Chen2010}. However, not until MMS was significant effort made to design sensors with synchronized timing with pre-launch end-to-end characterization of sensor performance intended to enable optimal merging of the in-flight magnetic field data \cite{Torbert2016,Fischer2016}. 
%\cite{Fischer2016} outline a merged magnetic field data product operating at 1024 Sa/S which is derived from 128 Hz fluxgate and 8192 Hz search-coil data with good signal to noise characteristics constructed from compensation filters.
The shared clock between the FIELDS SCM and MAGs, as well as their simultaneous continuous survey mode operation, likewise facilitates a merged SCM and MAG (SCaM) data product. In order to produce the merged SCaM data product, accurate representations for the complex frequency responses for the individual sensors are required. In addition to the individual characterization of the instruments, multiple efforts were made to inter-calibrate the sensors; however, no strict end-to-end calibration was performed as in \cite{Fischer2016}. Original ground testing was performed at the Acu\~{n}a Test facility using FIELDS engineering model hardware; subsequent testing was performed on flight model hardware during final stages of integration onto the PSP spacecraft, verifying the instrument gain and phase characteristics. In addition to the characterization of frequency response, the merged SCaM data product ideally attains minimal noise characteristics. Accordingly, an accurate description of the individual MAG and SCM sensor noise floors is necessary.

To provide an optimal signal to noise merging coefficients, the noise floors of each instrument are assumed to be incoherent, mean zero, gaussian processes. The spectral composition of the instrumental noise was determined during ground testing. The SCM sensitivity was characterized at the magnetic test facility in Chambon-la-For\^et. In addition to the internal sensor noise,  the DFB analog electronics as well as analog to digital conversion (quantization) of the SCM signal contribute to the end-to-end instrumental noise. The end-to-end noise floor of the MAGs, incorporating quantization and analog electronic noise, were determined in laboratory using measurements taken over several hours inside of a $\mu$-metal container.

The FIELDS SCaM merging procedure is designed to maintain an optimized signal to noise ratio. Each sensor observes the environmental field, which is a coherent signal between two sensors, in superposition with incoherent, zero mean noise. 

\begin{eqnarray}
{B_1}=B(t) +n_1(t)\\
{B_2}=B(t) +n_2(t)
\end{eqnarray}

The merged signal $B_m$ is given as a linear combination of the individual sensors, weighted by coefficients $\alpha_1$ and $\alpha_2$ which maintain an optimal signal to noise ratio.

As instrumental noise from each sensor has different spectral characteristics, we develop frequency dependent merging coefficients through consideration of the spectral representation of the linear combination of signal and noise terms 
\begin{eqnarray}
\tilde{B}_m(\omega)=\alpha_1\tilde{B}(\omega) +\alpha_2\tilde{B}(\omega) \\
\tilde{N}_m(\omega)=\sqrt{\alpha_1^2 \tilde{n_1}^2 +\alpha_2^2 \tilde{n_2}^2}\end{eqnarray}

where the merged noise $\tilde{N}_m(\omega)$ corresponds to the error of each signal, weighted and added in quadrature \cite{Kahn1954,Brennan1959}. The condition $\alpha_1 +\alpha_2 =1$ is required such that the merged signal is equal to the coherent environmental field observed by each sensor.

Because signal amplitudes are ideally equal in either sensor, optimizing the ratio $B_m/N_m$ leads to frequency dependent solutions for $\alpha_1(\omega)$ and $\alpha_2(\omega)$ which are independent of the environmental signal, and determined by the spectral composition of the noise floors:
\begin{eqnarray}\label{eq:alpha}
 \alpha_1(\omega)=\frac{\tilde{n_2}^2}{ \tilde{n_1}^2 +\tilde{n_2}^2}\\
\alpha_2(\omega)=\frac{\tilde{n_1}^2}{ \tilde{n_1}^2 +\tilde{n_2}^2}
\end{eqnarray}
where  $\tilde{n_1}^2$ and  $\tilde{n_2}^2$ are computed as the spectral densities of the instrument noise. For FIELDS, the coefficients $\alpha_{MAG}$ and $\alpha_{SCM}$ correspond to an effective weighting in instrumental gain which preserves an optimized signal to noise ratio for the merged SCaM data product. The SCM sensor coordinate system is not initially aligned with the MAG sensor axes, accordingly a rotation matrix $\bvec{R}$ is applied to bring the SCM measurements in sensor coordinates, $\bvec{B}_{SCM}'$, into alignment with the MAG coordinate system,
%
%\begin{eqnarray}
%\left(0.8165 & -0.4082 &-0.4082\\
%0.0000 & -0.7071 & 0.7071\\
%-0.57728773 & -0.57738150 & -0.57738158\right)
%\end{eqnarray}

  \begin{eqnarray}
  \bvec{R} =\left(\begin{array}{ccc}
      0.8165 & -0.4082 &-0.4082\\
      0.0000 & -0.7071 & 0.7071\\
-0.577 & -0.577& -0.577
      \end{array}\right)
   \end{eqnarray}

\begin{equation} \bvec{B}_{SCM}=\bvec{R}\cdot\bvec{B}_{SCM}'.\end{equation} Adhering to an optimal signal to noise merger, spectral composition of the noise of the rotated SCM vector time series in MAG sensor coordinates is then taken as the quadrature weighted error of the SCM sensor axis noise, assuming independence in each sensor channel: i.e. \begin{equation}n_{SCMx}^2(\omega)= R^2_{xx'}n^2_{SCMx'} +R^2_{xy'}n^2_{SCMy'}+R^2_{xz'}n^2_{SCMz'}.\end{equation} The MAG orthogonalization matrix and rotation from sensor to spacecraft coordinates is approximately equal to the identity matrix such that the measured noise floor for each sensor axis is used without contribution from the other axes.

\begin{figure}\centering
\includegraphics{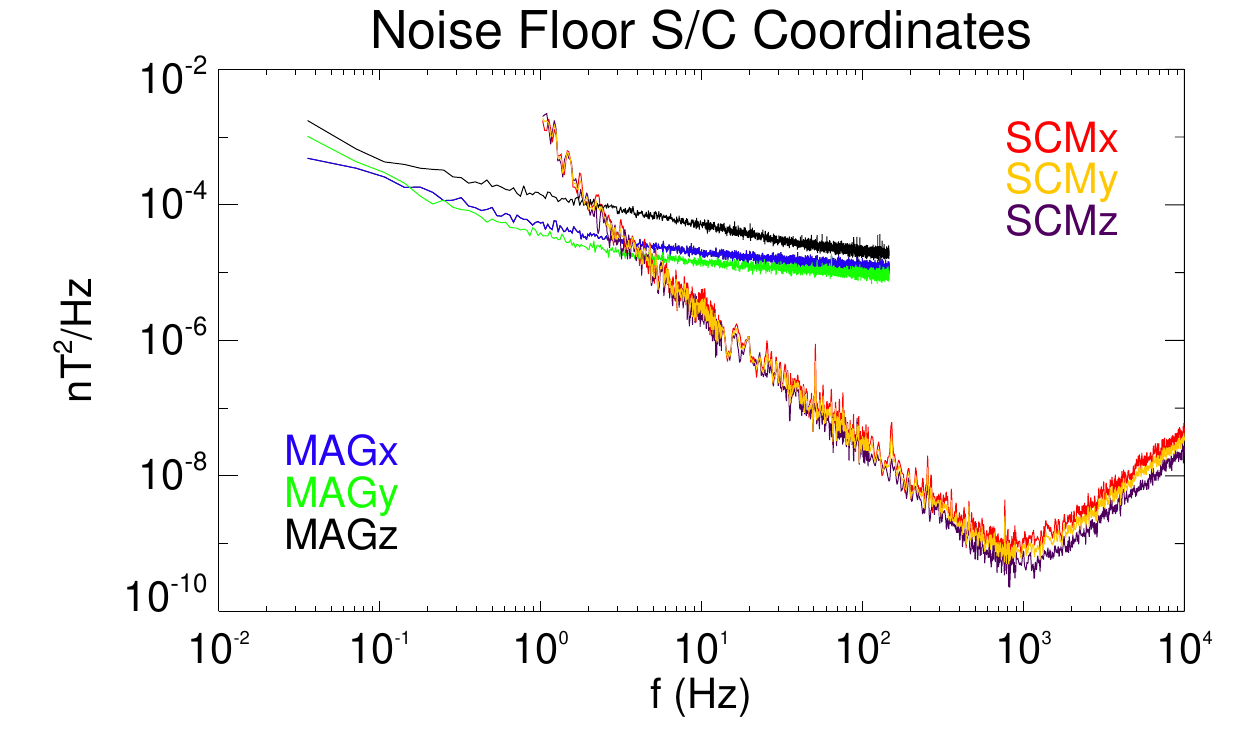}
\caption{Empirically determined noise floors of PSP FIELDS MAGo $x,y,z$ axes (blue, green, black) and the SCM low frequency $x, y, z$ axes with the DFB sampling in a high gain state (red, orange, purple) rotated into MAGo sensor coordinates.}
\label{merging_noise}
\end{figure}

\begin{figure}\centering
\includegraphics{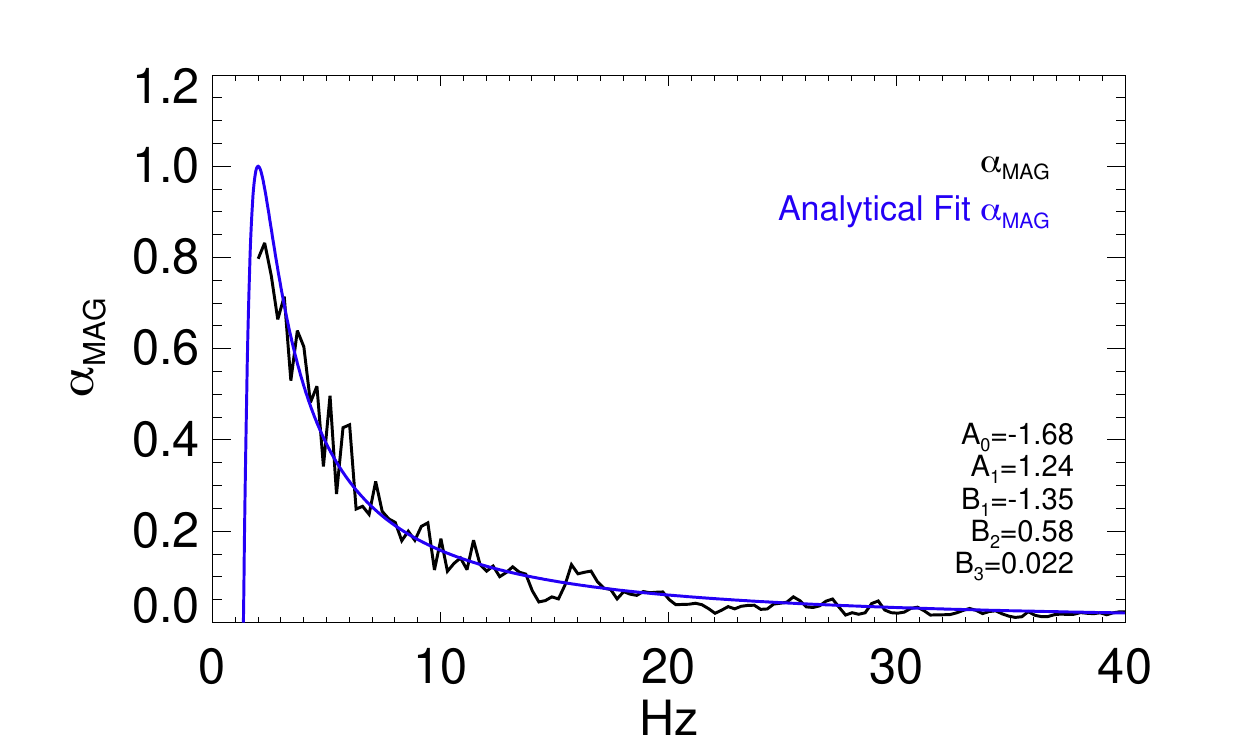}
\caption{Nonlinear least square fit of three pole, one zero rational function to $\alpha_{MAG}$ for $f>f_0$ with $f_0=2$ Hz. At 2 Hz, $\alpha_{MAG}$ is set to unity, the fit function is constrained to maintain a continuous value and first derivative, such that an extremum is obtained.}
\label{merging_fit}
\end{figure}

 Figure \ref{merging_noise} shows empirically determined noise-floors for both the SCM and MAGo associated with mean-zero stochastic fluctuations limiting each sensors sensitivity. Merging coefficients are obtained using Equation \ref{eq:alpha}. Since the
empirically determined noise spectrum is continuous, a smooth weighting of the merged signals is obtained by approximating the MAG merging coefficient using a real-valued rational function of the form \begin{equation}\label{eq:ratfit}\hat{\alpha}_{MAG}(f)=\frac{N(f)}{D(f)} = \frac{ \sum_0^n A_n f^n }{1 +\sum_1^m B_mf^m}.\end{equation}  Below $f_0=2$ Hz the sensitivity of the SCM drops significantly and the full MAG signal is used, i.e. $\alpha_{MAG} =1$ for $f\leq 2$ Hz. Above 2 Hz Equation \ref{eq:ratfit} is applied. The coefficients and order of the fit rational function are determined using non-linear least squares fitting \cite{Markwardt}. Constraints are imposed on $\hat{\alpha}_{MAG}(f)$ to ensure continuity such $\hat{\alpha}_{MAG}(f_0)=1$ and $\hat{\alpha}_{MAG}'(f_0) =0$, where $f_0$=2 Hz. Figure \ref{merging_fit} shows the $n=1, m=3$ (one zero, three pole) fit for $ \hat{\alpha}_{MAG}$.

\begin{figure}\centering
\includegraphics{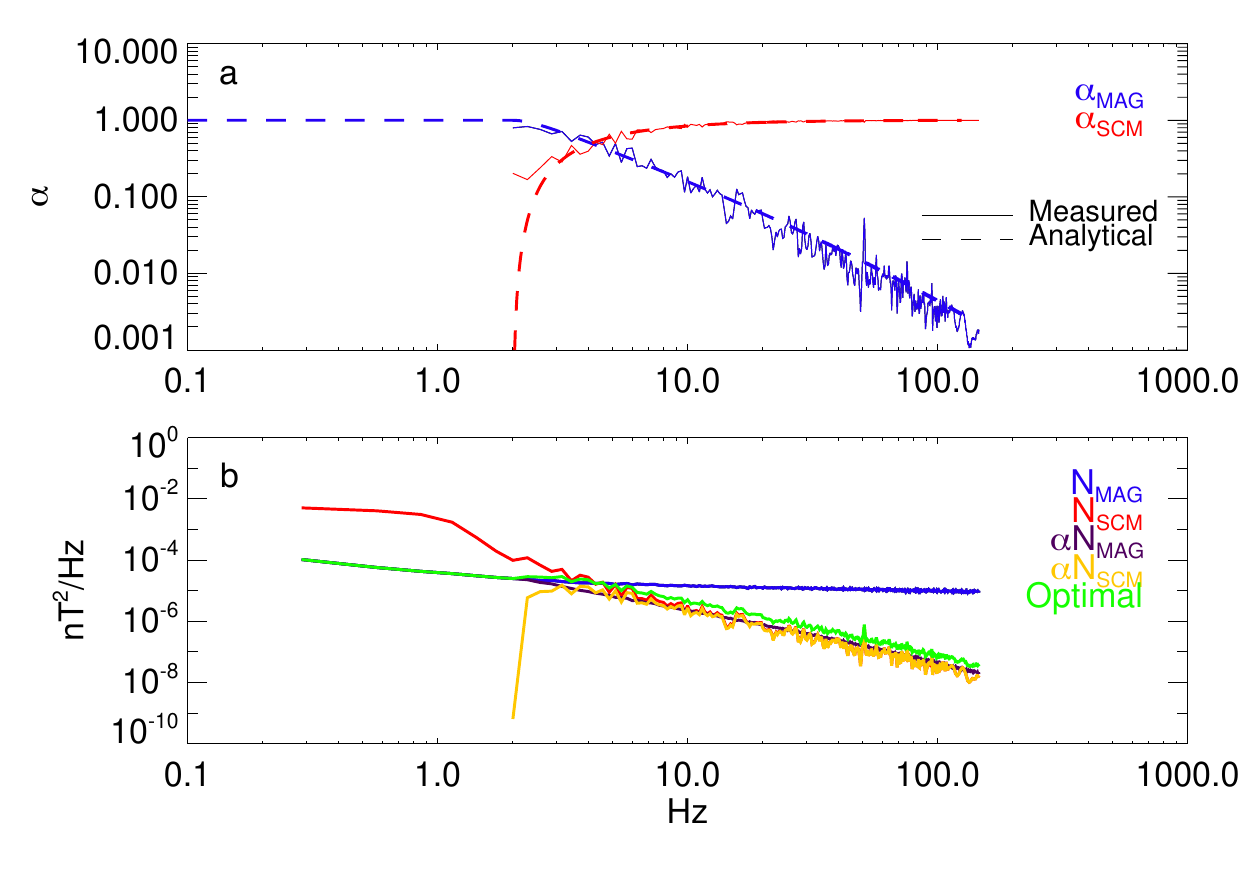}
\caption{(a) Analytical approximations of MAG and SCM merging coefficients ($y$-axis shown). (b) Noise floors of MAG (blue) and SCM (red), the coefficient weighted noise floors are shown for MAG (purple) and orange (SCM). The optimal noise floor is plotted in green.}
\label{merging_alpha}
\end{figure}

Fitting $\alpha_{MAG}(f)$ with boundary conditions at $f_0$=2 Hz decreases available degrees of two freedom such that rational functions with three or more fit parameters are required for a reasonable approximation. Figure \ref{merging_fit} shows the best fit rational function (Equation \ref{eq:ratfit}) with three poles and one zero (e.g. $m=3$ and $n=1$), to the MAG merging coefficient, $\alpha_{MAG}$ with applied boundary conditions at 2 Hz. Ensuring a piecewise continuous merging coefficient requires constraining the function at 2 Hz to unity gain and a local extremum. For $f< 2$ Hz, the coefficient $\alpha_{MAG}$ is explicitly set to unity, e.g. Figure \ref{merging_alpha}, while for $f > 2$ Hz, $\alpha_{MAG}=\hat{\alpha}_{MAG}$. The SCM coefficients are determined from ${{\alpha}}_{SCM} =1-{\alpha}_{MAG}$.  The optimal merging coefficients $\alpha_{MAG}$ and $\alpha_{SCM}$ are shown in Figure \ref{merging_alpha}(a). The the weighted instrumental noise floors, and the optimal merged SCaM noise floors are shown in Figure \ref{merging_alpha}(b).

%\begin{eqnarray}
%R'(f_0)=\frac{N'(f_0)D(f_0) -N(f_0)D'(f_0)}{D^2(f_0)}= \\
%\frac{N'(f_0) N(f_0)/R(f_0) - N(f_0)D'(f_0)}{N^2(f_0)/R^2(f_0)}=\\
%\frac{N'(f_0)R(f_0) - R^2(f_0)D'(f_0)}{N(f_0)}.
%\end{eqnarray}
%
%
%Representing the polynomials $N(f)$ and $D(f)$ as matrices
%\begin{eqnarray}
%N=\begin{pmatrix}
%A_0 & A_1 & \cdots &A_n
% \end{pmatrix} 
% \begin{pmatrix}
%1\\ x \\ \vdots \\x^n
% \end{pmatrix}
%\end{eqnarray}
%
% The derivative operator can be written as a matrix 
% \begin{equation}
% \Delta=
%  \begin{pmatrix}
%0& 1& 0&\cdots &0\\
%0& 0 & 2& \cdots &0\\
%0& 0& 0& \cdots &0\\
% \end{pmatrix}
% \end{equation}
% 
% such that 
% \begin{equation}N' = \left(\Delta  \begin{pmatrix}
%A_0 & A_1 & \cdots &A_n
% \end{pmatrix}^\dagger \right)^\dagger
% \begin{pmatrix}
%1\\ x \\ \vdots \\x^n
% \end{pmatrix}
% \end{equation}
% 
% Analogously writing D'(f) using the $\Delta$ provides an efficient method to determine solutions to the two linear equations
%\begin{eqnarray}
%R(f_0)D(f_0)=N(f_0)\\
%R'(f_0)N(f_0)=\Delta N(f_0)R(f_0) - R^2(f_0)\Delta D(f_0).
%\end{eqnarray}
%
%
\section{Calibration and Merger of In-Flight Survey Data}
\label{sec:inflight}

\begin{figure}\centering
\includegraphics{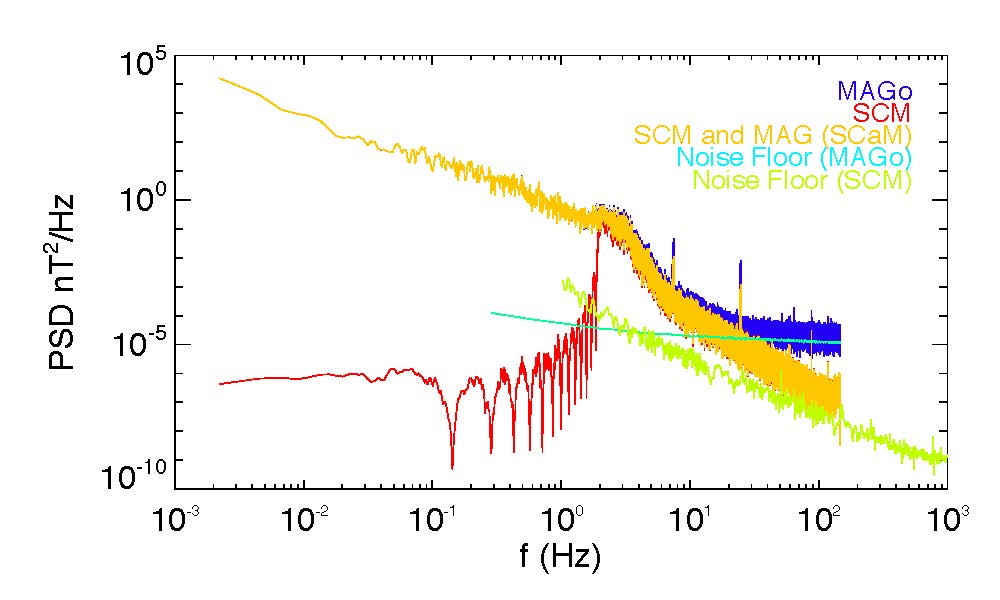}
\caption{Power spectra densities of observed magnetic field in the spacecraft coordinate $y$ direction from $\approx$ 1 hour interval (2018-11-05/00:00:-01:00) calculated with MAG (blue), SCM (red), and merged SCM and MAG (SCaM, orange), time series. Sensor noise floors are shown for the MAG (teal) and SCM (green).}
\label{merging_spect}
\end{figure}

\begin{figure}\centering

\includegraphics{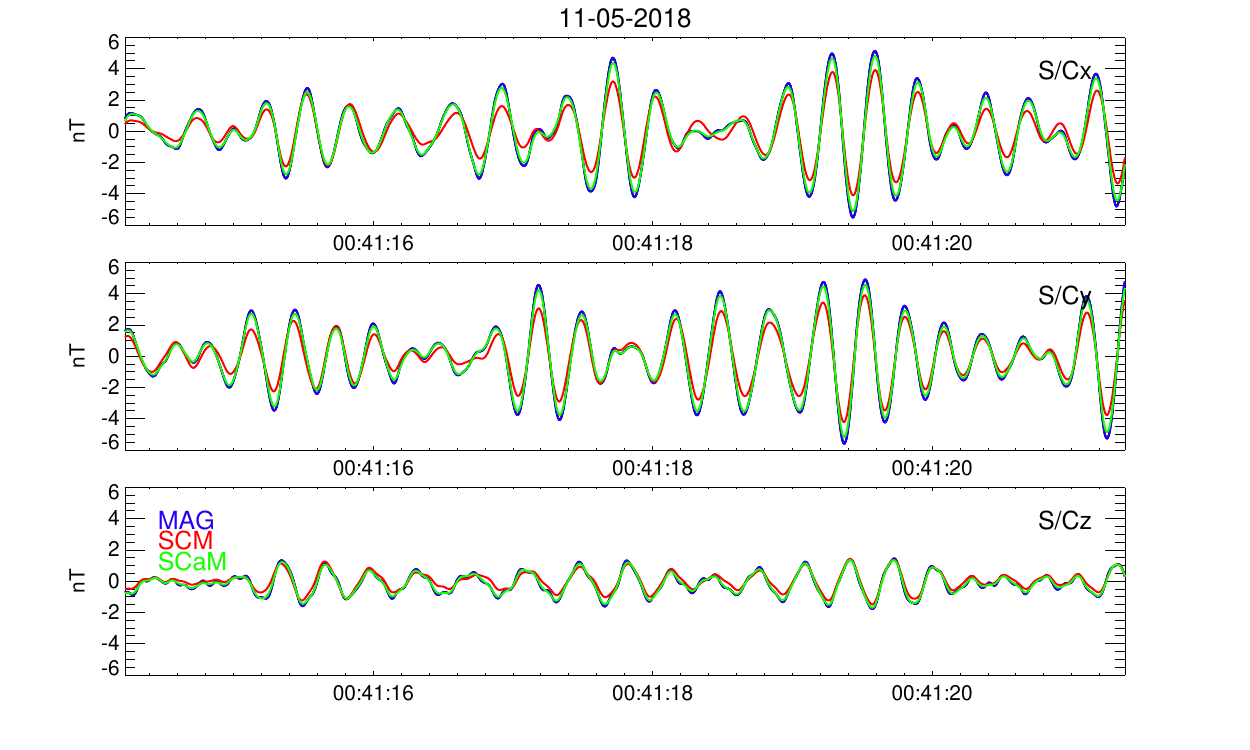}
\caption{Eight second bandpass filtered waveform near the merging crossover frequency (2-12 Hz) from PSP/FIELDS survey magnetic field data in S/C coordinates. Good phase and gain agreement is observed in between the MAG (blue) SCM (red) and merged SCaM (green) time series.}
\label{merging_waves}
\end{figure}

During the PSP perihelion encounters, survey mode data are acquired at different cadences, typically varying with solar distance, in order to balance science objectives with telemetry constraints. When possible the FIELDS team intends to operate the instruments with a single high data rate (292.969 Sa/s) over the entire perihelion encounter period. To date, the lowest cadence survey rate during the perihelion encounter is 73.24 Sa/s.

The calibration kernel, corresponding to the inverse response of the MAG instrument response is weighted by the appropriate merging gain coefficients, $\alpha_{MAG}$ to construct the contribution from MAGo to the merged time series. This weighted time-series, subsequently undergoes calibration processes associated with orthogonalization and spacecraft field removal used in generation of public un-merged data. The SCM is similarly calibrated using the instrumental response function with gain weighted by the merge coefficients $\alpha_{SCM}$; the gain and gain phase shifts associated with digitization by the FIELDS DFB are additionally corrected for \cite{Malaspina2016}. Once convolved with calibration kernels, the SCM is rotated into the MAGo coordinate system. 
%Time tags for each instrument are corrected using spacecraft SPICE kernels provided by PSP mission operations. 
The weighted MAG time series is then interpolated onto the SCM time abscissae. Interpolation onto the SCM time tags are used to preserve the high frequency component of the SCM without introducing artifacts associated with interpolation. The time series are directly summed to generate the merged data set with optimal signal to noise ratio. The merged SCaM data is considered a level 3 (L3) data product.

Power spectra from an approximate hour long interval ($2^{20}$ samples at 292.969 Sa/s) starting 11/05/2018T00:00 during the first PSP perihelion is highlighted in Figure \ref{merging_spect} to demonstrate the results of our calibration and merging algorithm. Power spectra for each of the MAG, SCM, and SCaM time series are computed as an ensemble average of eight power spectra of $2^{17}$ samples. Figure \ref{merging_spect} additionally shows noise-floors associated with the MAG and SCM instruments. Good agreement is observed between the merged data and spectra from either individual instrument. The in-flight observed noise-floor of the MAGo is consistent with on-ground measurements. The SCM noise floor performs similarly to preflight measurements; a slight increase in the sensitivity is observed relative to ground testing which is attributable to a decrease in thermal noise in the instrument. Broadband spectral features near the crossover frequency, corresponding to coherent wave features at several Hz, are captured by both the MAG and SCM and are thus useful in analyzing the performance of the SCaM merging algorithm \cite{Bale2019,Bowen2019}. Digital filters are applied to bandpass the MAG, SCM, and merged SCaM time series to between 2 and 12 Hz in order to directly compare the time series in the crossover bandwidth, without contribution from low or high frequency signals.  Figure \ref{merging_waves} shows excellent qualitative agreement in phase between the three different axes.
 
\begin{figure}\centering
\includegraphics{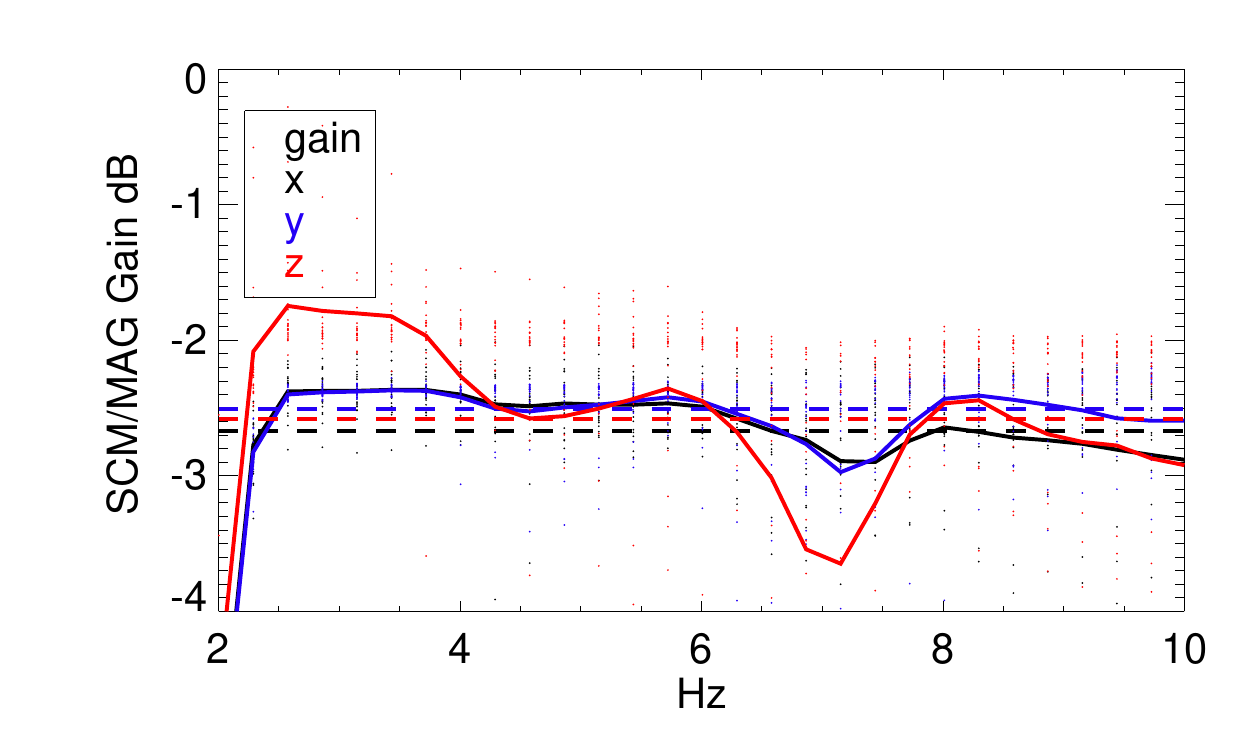}
\caption{Gain difference between SCM and MAG measured over their shared observations range ($x$, $y$, $z$ axes shown in black, blue, red). The measured distribution of gain differences is plotted as a set of points. The mean at each frequency is plotted as a sold line, while the median gain difference in each axis from 3-10 Hz is plotted as a dashed line. The average gain difference is roughly constant over this range. The feature at 7 Hz corresponds to a reaction wheel (which has a lower signal in the SCM due to the relative positioning of the sensors).}
\label{gain_error}
\end{figure}

However, in order to ensure quality of the merged SCaM data product, the calibrated, but un-merged, MAG and SCM observations must be analyzed to verify the conditions necessary for the weighted-gain merging algorithm developed in Section \ref{sec:merging}: i.e. the MAG and SCM cross calibration, including time synchronization and gain matching, must be verified. Careful inspection of Figure \label{merging_waves} shows a small gain discrepancy between the MAG and SCM amplitudes. Analysis of the gain discrepancy is required to ensure an artifact-free merged data product. Quantitative determination of the relative MAG and SCM gain calibrations is performed by separating the full day of encounter data from Nov 05, 2019 into 22 non-overlapping intervals of $2^{20}$ samples (a one hour interval where the SCM was in a low-gain state was omitted). The vector spectral density for each interval is estimated for both MAG and SCM sensors by ensemble averaging the power spectrum of 1024 non-overlapping sub intervals computed via FFT e.g.: $$S_{MAG(f)}=\langle\mathcal{F}\{B_{MAG}(t)\}\mathcal{F}^{\dagger}\{B_{MAG}(t)\}\rangle,$$ where $\mathcal{F}\{...\}$ is the Fourier transform and $\langle ...\rangle$ denotes ensemble averaging; a Blackman-Harris window is used to prevent spectral leakage. The frequency dependent gain is then obtained as 
$$G(f)=10\mathrm{Log}_{10} \frac{S_{SCM}}{S_{MAG}}.$$ 
Figure \ref{gain_error} shows the measured distribution of $G(f)$ for each vector component as well as the mean at each frequency, and the median gain error computed between 3 and 10 Hz. Systematic gain differences are measured  in the $x$, $y$, and $z$ directions of -2.67, -2.50, and -2.58 dB.  These values indicate that the typical SCM amplitude is approximately $75\%$ of the measured MAG signal. 

Due to the relatively stable gain discrepancy in frequency and time, the SCM may be gain-matched to the MAG through multiplication of scaling factors (1.36,1.33, and 1.34) for the respective $x$, $y$,and $z$ axes. This correction is required in-order to remove artifacts associated with merging signals with un-equal amplitudes. The difference between on-ground and in-flight gain measurements are likely due to differences in the SCM operating temperature and small discrepancies caused by the
matching between SCM and DFB, which unfortunately were not quantified due to lack of end-to-end calibration. Continued efforts to quantify and monitor variations in the gain-matching coefficients will be performed throughout the mission. Additionally, both gain-matched and nominal calibrations will be available for public use; though the authors stress that use of non-gain matched data may lead to artifacts in the transition between the MAG and SCM sensor ranges.

A quantitative determination of the accuracy of the timing between the MAG and SCM data is performed by computing the short time Fourier transform cross-spectra of each of the three combinations of signals. The short time cross spectra is defined as
\begin{equation}
S_{12}=\langle\mathcal{F}\{B_1(t)\}\mathcal{F}^{\dagger}\{B_2(t)\}\rangle
\end{equation} 

\begin{figure}\centering
\includegraphics{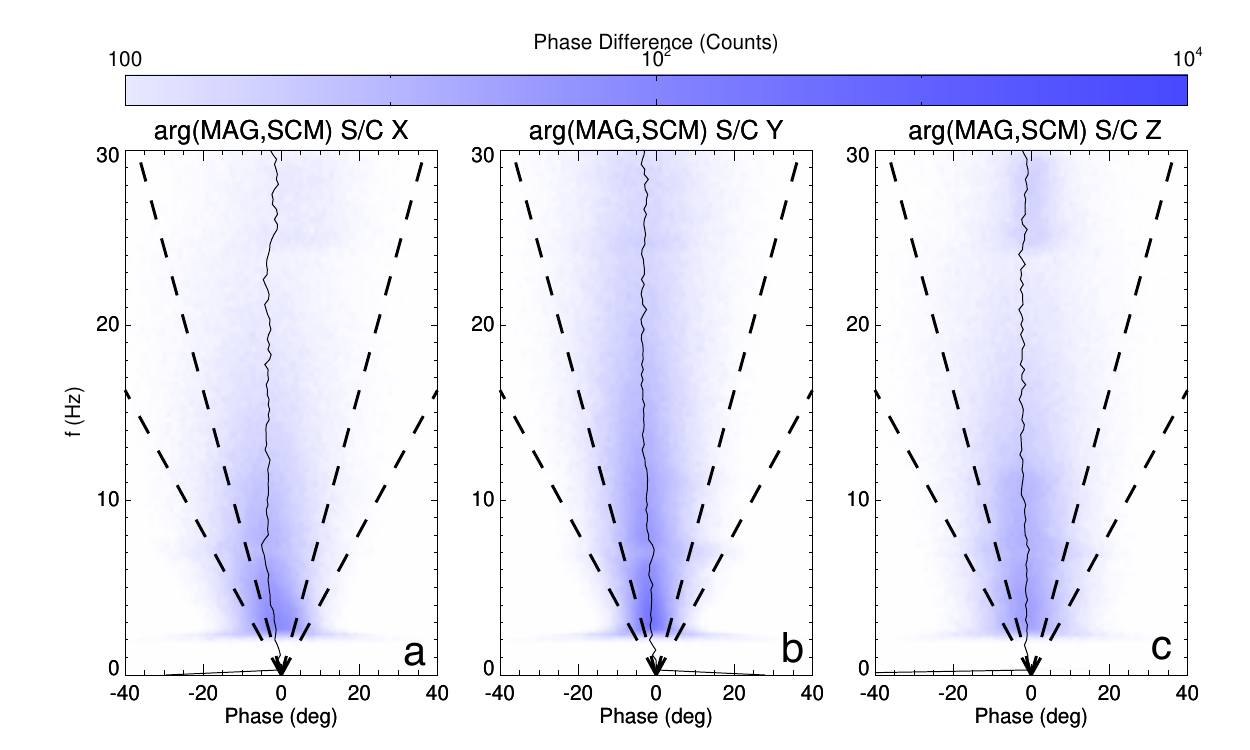}
\caption{Distribution of measured phase delays between the MAG and SCM in spacecraft coordinates ($x$, $y$, $z$) as a function of frequency shown respectively in panels (a ,b, c). The solid black line shows the mean phase error at each frequency. The dashed  black lines show linear phase error associated with one and two sample periods ($\Delta t \sim 340 \mu$s).}
\label{merging_calphase}
\end{figure}

The argument of the cross spectra gives the phase delay between the two signals at a given frequency ${\textrm{arg}}(S_{12})=\textrm{tan}^{-1}\left(\frac{\textrm{Im}\{S_{12})\}}{\textrm{Re}\{S_{12}\}}\right)$. As each sensor observes the same time series, zero-phase difference should exist at each frequency between the sensors. Each of the 22 intervals on 11/05/2018 used in gain calibration are separated into 1024 sub-intervals. The sub-division allows for the calculation of 22528 individual cross-spectra.  The distribution of phase delay as a function of frequency is then calculated between the MAG and SCM using the ensemble of cross-spectra.

Figure \ref{merging_calphase} shows the measured distributions of phase difference between the MAG and SCM obtained via cross spectra. The MAG and SCM are shown to be in good agreement in the cross over frequencies: at 4 Hz the mean time-delay between the MAG and SCM measurements is 190 $\mu$s, 84 $\mu$s,  and 100 $\mu$s for the respective $x$,$y$, and $z$ axes; the standard deviations are 76 $\mu$s, 59 $\mu$s, and 89$\mu$s respectively. For each vector component, approximately two standard deviations of the measured ensemble fall within $1 \Delta t$ ($\sim 340 \mu$s). These results show that the phase alignment of the MAG and SCM is accurate to within a small phase error in the frequencies surrounding the cross-over point.

Analysis of the relative phase between the SCM and MAG observations verifies timing accuracy to within a single sample period. Additionally, quantification of the relative gain between the instruments allows for the empirical matching of the SCM signal to MAG levels such that a smooth transition over the sensor cross-over range is obtained. Establishing L3 calibrations for the MAG and SCM provide time-synchronized and gain matched signals such that the direct sum of the signals, with gains weighted by $\alpha_{MAG}$ and $\alpha_{SCM}$, results in an optimal signal-to-noise merged SCaM data product .

%\begin{figure}\centering
%\includegraphics{fig9}
%\caption{(a) Distribution of phase delays between the MAG and SCaM data product as a function of frequency. (b) Distribution of phase delays between the SCM and SCaM data product as a function of frequency}
%\label{merging_mgd}
%\end{figure}

\subsection{Merging DFB Burst Data}
\begin{figure}\centering
\includegraphics{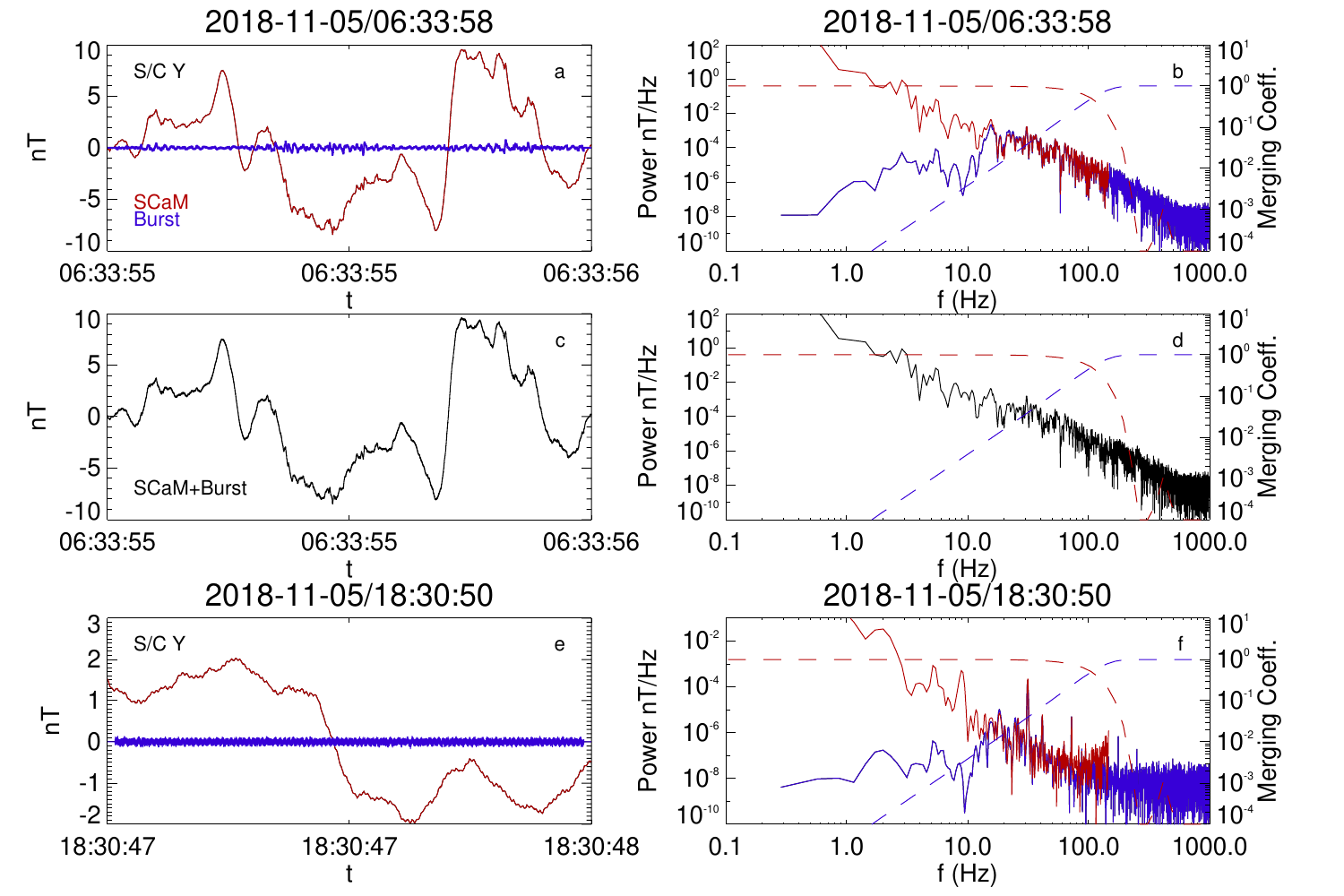}
\caption{(a) Burst waveform from FIELDS DFB on 2018-11-05/06:33:58 (blue) with simultaneous SCaM (red) waveform data. (b) Power spectral density from burst interval on 2018-11-05/06:33:58 dashed lines show the corresponding merging coefficients for the DFB burst (blue) and SCaM (red). (c) Merged DFB and SCaM waveform. (d) Merged power spectral density. (e) Waveform data from DFB burst on 2018-11-05/06:33:58 with SCaM data. (f) Power spectral density from 2018-11-05/06:33:58 interval, spectral flattening in the survey waveform can occur either when the DFB quantization limit is reached, or when narrowband spectral features are present above survey Nyquist frequency.
}
\label{BurstCal}
\end{figure}
In addition to survey waveform data, the FIELDS DFB  produces high resolution burst data from SCM measurements at a maximum sample rate of $f_{brst}=150$ kSa/s. For the three low frequency SCM windings, this is significantly higher than the instrumental 3 dB roll off at $\sim 17$ kHz. The burst buffer is taken as a $N_{brst}=2^{19}$ sample waveform lasting $\sim 3.5$ seconds. Data from the SCaM product is combined with the DFB burst measurements to provide the low frequency spectral composition to contextualize the DFB bursts. The SCM transfer function is applied to the burst data using a finite impulse response calibration kernel of $M_{brst}=$16384 filter coefficients (taps); a cutoff is applied at the high frequency SCM 3 dB roll-off ( $\sim$ 17 kHz) to prevent amplification of high frequency noise. Frequencies $f< f_{brst}/M_{brst}$ (e.g.  $\sim 10$ Hz for $M_{brst}=16384$ and $f_{brst}=150$ kSa/s) cannot be captured using a convolution kernel; however, the merged SCaM data is optimized to provide high signal to noise measurements in this frequency range. 

Figure \ref{BurstCal}(a) demonstrates a comparison of the SCaM and DFB burst waveforms from a burst on 2018-11-05/06:33:58. Figure \ref{BurstCal}(b) shows power spectral densities for the interval. To combine low frequency spectral components with the calibrated burst data, the SCaM data is interpolated onto the DFB burst time-tags, which is rotated into the S/C coordinate system. At frequencies above $\sim 10$ Hz the SCaM data is predominantly derived from the SCM and the intrinsic noise of the DFB burst and SCaM data are thus identical; however, correcting the attenuation from the DFB anti-aliasing filters during the SCaM calibration results in the amplification of noise in the high frequency end of the survey wave-form data. The difference in noise level between the burst and SCaM data corresponds precisely to the DFB anti-aliasing filter, which is taken as the weighting coefficients, shown Figure \ref{BurstCal}(b,d,e) to merge the SCaM and burst data. Figure \ref{BurstCal}(c) shows the merged SCaM data with DFB burst waveform data; the corresponding merged power spectral density is depicted in Figure \ref{BurstCal}(d).

A second burst interval from 2018-11-05/18:30:50 is shown in Figure \ref{BurstCal}(e-f). On occasion, spectral flattening is observed in the high frequency component of the SCM and SCaM data, e.g. Figure \ref{BurstCal}(f). Using DFB burst waveform data, it has been determined that this effect likely occurs due to the presence of relatively large amplitude narrowband spectral noise located immediately above the survey waveform Nyquist frequency. Such flattening is also evident with the noise floor of the SCM is reached. Ongoing efforts are made to characterize narrowband spectral noise and its effect on magnetic field measurements made by FIELDS \cite{Bowen2019}.

\subsection{In-Flight Issues with SCMx Axis}

Since March 2019, the low-frequency $B_x$ channel of the SCM has deviated from nominal operation whenever the sensor is shaded by the TPS, e.g. during perihelion encounters. The main symptoms of this anomaly are a much higher sensitivity to periodic current surges in the SCM heater and a drop in sensitivity in the low frequency wave-form channels. This drop is equivalent to the response of an additional 1st order high-pass filter with a cutoff at 1 kHz. This sudden change in sensitivity mostly impacts measurements of $B_x$ below approximately 600 Hz. The three components of the SCM are rotated in the frame of the MAG before merging. Accordingly the anomaly in one single channel will affect all three vector components in the S/C spacecraft coordinates, making it impossible to properly merge the signals from the two instruments. By rotating the MAG measurements into the SCM frame, it is possible to merge two components. For PSP's first encounter the full vector merged product in spacecraft and RTN coordinates will be produced. For later events, a 2D merged product will be distributed using the SCM y and z axes. Consequently, the MAG is the only remaining 3-axis measurement below approximately 150 Hz; while the SCM provides vector measurements above approximately 600 Hz. The degraded phase and amplitude response of the SCM maintains remarkable stability, which suggests that it remains possible to deliver properly calibrated data outside of the intermediate frequency range.

\section{Conclusion}

This manuscript reports on the development, implementation, and performance of an algorithm to merge magnetic field observations from the PSP/FIELDS  fluxgate (MAG) and search-coil (SCM) to create a merged SCM and MAG (SCaM) data product. The techniques used for PSP/FIELDS are similar to efforts made by \citeA{Fischer2016} to combing magnetic field measurements from instrumentation on MMS, which attempt to maintain optimal signal to noise characteristics. These merging algorithms have heritage from techniques developed in radio-systems engineering for linear diversity combining \cite{Kahn1954,Brennan1959}. The optimal merging methods takes into account sensor design and operation of the MAG and SCM instrumentation, in order to construct a merged data product with spectral composition which smoothly transitions between the MAG at low frequencies and the SCM at high frequencies with a cross over between $\sim3-10$ Hz.  In the cross-over range of frequencies, both sensors contribute significantly to the merged SCaM data product. Using in flight analysis of the calibrated FIELDS observations, we demonstrate that the MAG and SCM sensors are in systematic agreement to within a small fraction of a sample period ( $<340\mu$ s), enabling a smooth transition between dominant signal in the cross over range without phase distortion of the measured waveforms. A small deviation in gain ($\sim$ 2 dB), likely due to temperature effects,  between the sensors is measured, which impacts the merging procedure and requires ongoing analysis and correction.

Additionally, the merging algorithm presented is used to combine burst data from the SCM, acquired by the FIELDS DFB at a 150 kSa/s sample rate, with survey rate data from the MAG and SCM at lower frequencies. The merged DFB burst data allows for analysis of magnetic field signals from DC to the SCM LF cutoff (17 kHz) within a single dataset. The successful merging of SCM survey rate data with DFB burst data is promising for ongoing efforts to merge burst measurements from the FIELDS TDS at 1.92 MSa/s with these lower frequency data products. Additionally, the algorithm outlined to merge magnetic field measurements serves as a starting point to merge survey and burst measurements of waveforms made by the FIELDS electric fields antennas as well as other timeseries.

In order to maintain the level three SCaM data product for continued public use, the FIELDS team intends to regularly update the merging algorithm using measured onboard frequency responses, temperatures, noise floors, MAG offsets etc. 
\bibliography{references.bib}
%%\bibliographystyle{natbib}
%\end{document}
\end{document}